\documentclass[a4paper,10pt]{article}
\usepackage{etex}

\usepackage{authblk}

\usepackage{lmodern}\usepackage[T1]{fontenc} 
\usepackage[utf8]{inputenc}
\usepackage[UKenglish]{babel}

\usepackage{amsmath, amssymb, mathtools}
\usepackage{hyperref}
\usepackage{booktabs}
\usepackage{subcaption} 
\usepackage{cprotect}

\usepackage[nounderscore]{syntax}
\usepackage{mod}
\usepackage{chemfig}

\usepackage{textcomp} 
\usepackage{listings}
\definecolor{purple2}{RGB}{153,0,153} 
\definecolor{green2}{RGB}{0,153,0} 
\lstdefinelanguage{pymod}{
  morekeywords={
    access,and,break,class,continue,def,del,elif,else,except,exec,finally,for,
    from,global,if,import,in,is,lambda,not,	or,pass,raise,return,try,while,
    False,True
  },
  morekeywords=[2]{
    abs,all,any,basestring,bin,bool,bytearray,callable,chr,
    classmethod,cmp,compile,complex,delattr,dict,dir,divmod,enumerate,eval,
    execfile,file,filter,float,format,frozenset,getattr,globals,hasattr,hash,
    help,hex,id,input,int,isinstance,issubclass,iter,len,list,locals,long,map,
    max,memoryview,min,next,object,oct,open,ord,pow,property,range,
    raw_input,reduce,reload,repr,reversed,round,set,setattr,slice,sorted,
    staticmethod,str,sum,super,tuple,type,unichr,unicode,vars,xrange,zip,apply,
    buffer,coerce,intern
  },
  sensitive=true,%
  morecomment=[l]\#,%
  morestring=[b]',%
  morestring=[b]",%
  morecomment=[s]{'''}{'''},
  morecomment=[s]{"""}{"""},
  morestring=[s]{r'}{'},
  morestring=[s]{r"}{"},%
  morestring=[s]{r'''}{'''},%
  morestring=[s]{r"""}{"""},%
  morestring=[s]{u'}{'},
  morestring=[s]{u"}{"},%
  morestring=[s]{u'''}{'''},%
  morestring=[s]{u"""}{"""},%
  basicstyle=\normalsize\ttfamily,   
  keywordstyle=\color{orange},       
  keywordstyle={[2]\color{purple2}}, 
  stringstyle=\color{green2},
  commentstyle=\color{red},
  showstringspaces=false,
  upquote=true,                      
}[keywords,comments,strings]
\lstdefinelanguage{modtex}[LaTeX]{tex}{
  keywordstyle=\color{blue},
  texcsstyle=*keywordstyle,
  morekeywords={graphGML,smiles,ruleGML,dpoRule},
  morekeywords={collapse,hydrogens,edges,as,bonds}
}

\newsavebox{\listingsBox}

\newcommand\derivation[3][]{%
  \begin{tikzpicture}[gCat]
    \node[vCat,label=below:$D$](D)	       {\includegraphics[#1]{#3.derD}};
    \node[vCat,label=below:$G$](G)[left=of D]  {\includegraphics[#1]{#3.derG}};
    \node[vCat,label=below:$H$](H)[right=of D] {\includegraphics[#1]{#3.derH}};
    \draw[eMorphism](D) to (G);
    \draw[eMorphism](D) to (H);
    \node[vCat,label=above:$L$](L)[above=of G] {\includegraphics[#1]{#2.derL}};
    \node[vCat,label=above:$K$](K)[above=of D] {\includegraphics[#1]{#2.derK}};
    \node[vCat,label=above:$R$](R)[above=of H] {\includegraphics[#1]{#2.derR}};
    \draw[eMorphism](K) to (L);
    \draw[eMorphism](K) to (R);
    \draw[eMorphism](L) to (G);
    \draw[eMorphism](K) to (D);
    \draw[eMorphism](R) to (H);
    \end{tikzpicture}%
}

\newcommand{\stratStyle}[1]{\texttt{\textbf{#1}}}

\DeclareMathOperator{\stratParallel}{\stratStyle{parallel}}

\usepackage[textsize=scriptsize, colorinlistoftodos]{todonotes} 

\newcommand{\mset}[1]{\left\{\!\!\left\{#1\right\}\!\!\right\}}

\newcommand{\rc}{\bullet}
\newcommand{\rcParallel}{\rc_\emptyset}
\newcommand{\rcFull}{\rc_\supseteq}
\newcommand{\rcFullRev}{\rc_\subseteq}
\newcommand{\rcPartial}{\rc_\supseteq^c}
\newcommand{\rcPartialRev}{\rc_\subseteq^c}
\newcommand{\rcCommon}{\rc_\cap}

\DeclareMathOperator{\img}{img}

\usepackage{marvosym}
\newcommand\corr{\text{ (\Letter)}}
\newcommand\email[1]{\texttt{#1}}

\title{
A Software Package for Chemically Inspired Graph Transformation
}
\author[1,9]{Jakob L.\ Andersen\corr}
\author[2,8]{Christoph Flamm}
\author[1]{ Daniel Merkle\corr}
\author[2-7]{ Peter F.\ Stadler}
\affil[1]{
    Department of Mathematics and Computer Science, University of Southern Denmark, Odense M DK-5230, Denmark\\
    \email{\{jlandersen,daniel\}@imada.sdu.dk}
}
\affil[2]{
    Institute for Theoretical Chemistry, University of Vienna, Wien A-1090, Austria\\
    \email{xtof@tbi.univie.ac.at}
}
\affil[3]{
    Bioinformatics Group, Department of Computer Science,and  Interdisciplinary Center for   Bioinformatics, University of Leipzig, Leipzig D-04107, Germany\\
    \email{stadler@bioinf.uni-leipzig.de}
}
\affil[4]{
    Max Planck Institute for Mathematics in the Sciences, Leipzig D-04103, Germany
}
\affil[5]{
    Fraunhofer Institute for Cell Therapy and Immunology, Leipzig D-04103, Germany
}
\affil[6]{
    Center for non-coding RNA in Technology and Health,  University of Copenhagen, Frederiksberg C DK-1870, Denmark
}
\affil[7]{
    Santa Fe Institute, 1399 Hyde Park Rd, Santa Fe NM 87501, USA
}
\affil[8]{
    Research Network Chemistry Meets Microbiology, University of Vienna,
    Wien A-1090, Austria
}
\affil[9]{
    Earth-Life Science Institute, Tokyo Institute of Technology, Tokyo 152-8550, Japan\\
    \email{jlandersen@elsi.jp}
}

\date{}

\usepackage{hyperref}
\begin{document}
\maketitle

\begin{abstract}
  Chemical reaction networks can be automatically generated from graph
  grammar descriptions, where transformation rules model reaction patterns.
  Because a molecule graph is connected and reactions in general involve
  multiple molecules, the transformation must be performed on multisets of
  graphs.  We present a general software package for this type of graph
  transformation system, which can be used for modelling chemical systems.  The
  package contains a C++ library with algorithms for working with
  transformation rules in the Double Pushout formalism, e.g., composition
  of rules and a domain specific language for programming graph language
  generation.  A Python interface makes these features easily accessible.
  The package also has extensive procedures for automatically visualising
  not only graphs and transformation rules, but also Double Pushout diagrams and
  graph languages in form of directed hypergraphs.  The software is
  available as an open source package, and interactive examples can be
  found on the accompanying webpage.
\end{abstract}

\section{Introduction}
It has been common practice in chemistry for more than a century to
represent molecules as labelled graphs, with vertices representing atoms, and
edges representing the chemical bonds between them \cite{sylvester}. It is
natural, therefore, to formalize chemical reactions as graph
transformations \cite{benko,Rossello:2004,Ehrig:2006,Flamm:10b}.  Many
computational tools for graph transformation have been developed; some of
them are either specific to chemistry \cite{Rossello:2005b} or at least
provide special features for chemical systems \cite{ggl}. General graph
transformation tools, such as AGG \cite{Taentzer:2004}, have also been used to
modelling chemical systems \cite{Ehrig:2006}.

Chemical graph transformation, however, differs in one crucial aspect from
the usual setup in the graph transformation literature, where a single
(usually connected) graph is rewritten, thus yielding a graph language.
Chemical reactions in general involve multiple molecules. Chemical graph
transformations therefore operate on \emph{multisets} of graphs to produce
a chemical ``space'' or ``universe''. 
A similar viewpoint was presented in \cite{Kreowski2011}, but here we let
the basic graphs remain connected, and multisets of them are therefore dynamically constructed and taken apart in direct derivations.

Graph languages can be infinite. This is of course also true for chemical
universes (which in general contain classical graph languages as subsets).
In the case of chemistry, the best known infinite universes comprise
polymers. The combinatorics of graphs makes is impossible in most cases to
explore graph languages or chemical universes by means of a simple
breadth-first search. This limitation can be overcome at least in part with
the help of strategy languages that guide the rule applications. One such
language has been developed for rewriting port graphs \cite{Kirchner11},
implemented in the PORGY tool \cite{Andrei11}. We have in previous work
presented a similar strategy language \cite{dgStrat} for transformation of
multisets of graphs, which is based on partial application of
transformation rules~\cite{ruleComp}.

Here, we present the first part of the software package \modName{} (in
short: \modAbbr{}), that contains a chemically inspired graph
transformation system, based on the Double Pushout formalism
\cite{handbook1}.  It includes generic algorithms for composing
transformation rules \cite{ruleComp}. This feature can be used, e.g., to
abstract reaction mechanisms, or whole pathways, into overall rules
\cite{trace}. \modAbbr{} also implements the strategy language
\cite{dgStrat} mentioned above. It facilitates the efficient generation of
vast reaction networks under global constraints on the system.  The
underlying transformation system is not constrained to chemical
systems. The package contains specialized functionalities for applications
in chemistry, such as the capability to load graphs from SMILES strings
\cite{smiles}.  This first version of \modAbbr{} thus provides the main
features of a chemical graph transformation system as described in
\cite{Yadav:2004}.

The core of the package is a C++11 library that in turn makes use
of the Boost Graph Library \cite{bgl} to implement standard graph
algorithms. Easy access to the library is provided by means of extensive
Python bindings. In the following we use these to demonstrate the
functionality of the package. The Python module provides additional
features, such as embedded domain-specific languages for rule composition,
and for exploration strategies. The package also provides comprehensive
functionality for automatically visualising graphs, rules, Double Pushout
diagrams, and hypergraphs of graph derivations, i.e., reaction networks.  A
\LaTeX{} package is additionally included that provides an easy mechanism
for including visualisations directly in documents.

In Section~\ref{sec:formalism} we first describe formal background for
transforming multisets of graphs.  Section~\ref{sec:graphsRules} gives
examples of how graph and rule objects can be used, e.g., to find morphisms
with the help of the VF2 algorithms \cite{vf2:1,vf2:2}.
Section~\ref{sec:ruleComp} and \ref{sec:dgStrat} describes the interfaces
for respectively rule composition and the strategy language.
Section~\ref{sec:figgen}, finally, gives examples of the customisable
figure generation functionality of the package, including the \LaTeX{}
package.

The source code of \modName{} as well as additional usage examples can be
found at \url{http://mod.imada.sdu.dk}.  A live version of the
software can be accessed at
\url{http://mod.imada.sdu.dk/playground.html}. This site also provides
access to the large collection of examples.

\section{Transformation of Multisets of Graphs}
\label{sec:formalism}

The graph transformation formalism we use is a variant of the Double Pushout (DPO)
approach (e.g., see \cite{handbook1} more details).
Given a category of graphs $\mathcal{C}$, a DPO rule is defined as a span $p = (L\xleftarrow{l}
K\xrightarrow{r} R)$, where we call the graphs $L$, $K$, and $R$
respectively the \emph{left side}, \emph{context}, and \emph{right side} of
the rule.  A rule can be applied to a graph $G$ using a match morphism
$m\colon L\rightarrow G$ when
the \emph{dangling condition} and the \emph{identification condition} are satisfied
\cite{handbook1}.  This results in a new graph $H$, where the copy of $L$
has been replaced with a copy of $R$.  We write such a direct derivation as
$G\xRightarrow{p, m} H$, or simply as $G\xRightarrow{p} H$ or $G\Rightarrow
H$ when the match or rule is unimportant. The graph transformation thus
works in a category $\mathcal{C}$ of possibly disconnected graphs.

Let $\mathcal{C}'$ be the subcategory of $\mathcal{C}$ restricted to
connected graphs. A graph $G\in\mathcal{C}$ will be identified with the
multiset of its connected components. We use double curly brackets
$\mset{\dots}$ to denote the construction of multisets. Hence we write
$G=\mset{g_1,g_2,\dots g_k}$ for an arbitrary graph $G\in\mathcal{C}$
with not necessarily distinct connected components $g_i\in\mathcal{C}'$.
For a set $\mathcal{G}\subseteq\mathcal{C}'$ of connected graphs
and a graph $G = \mset{g_1, g_2, \dots, g_k}\in \mathcal{C}$
we write $G\in^* \mathcal{G}$ whenever $g_i\in\mathcal{G}$ for all $i=1, \dots, k$.

We define a graph grammar $\Gamma(\mathcal{G},\mathcal{P})$
by a set of connected starting graphs $\mathcal{G}\subseteq\mathcal{C}'$,
and a set of DPO rules $\mathcal{P}$ based on the category $\mathcal{C}$.
The language of the grammar $L(\mathcal{G},\mathcal{P})$ includes the starting graphs $\mathcal{G}$.
Additional graphs in the language are constructed by iteratively finding direct derivations $G\xRightarrow{p} H$ with $p\in \mathcal{P}$ and $G, H\in\mathcal{C}$ such that $G\in^*L(\mathcal{G},\mathcal{P})$.
Each graph $h\in H$ is then defined to be in the language as well.
A concise constructive definition of the language is thus
$L(\mathcal{G},\mathcal{P}) = \bigcup_{k=1}^{\infty} \mathcal{G}_k$ with
$\mathcal{G}_1 = \mathcal{G}$ and
\begin{align*}
  \mathcal{G}_{k+1} &= \mathcal{G}_k \cup \bigcup_{p\in \mathcal{P}}
  \{h\in H \mid \exists G \in^* \mathcal{G}_k: G\xRightarrow{p} H\}
\end{align*}

In \modAbbr{} the objects of the category $\mathcal{C}$ are all undirected
graphs without parallel edges and loops, and labelled on vertices and edges
with text strings.  The core algorithms can however be specialised for
other label types.
We also restrict the class of morphisms in $\mathcal{C}$ to be injective,
i.e., they are restricted to graph monomorphisms.
Note that this restriction implies that the identification condition of rule application is always fulfilled.

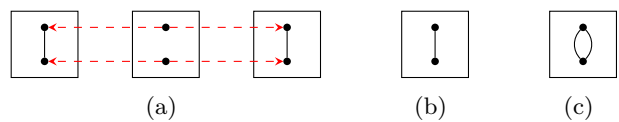
\begin{figure}[tbp]
\centering
\subcaptionbox{\label{fig:pushout:simple:span}}{
\begin{tikzpicture}[gCat, remember picture]
\node[vCat] (C) {
\begin{tikzpicture}[gDot, remember picture]
\node[vDot] (v-C-1) {};
\node[vDot, below=of v-C-1] (v-C-2) {};
\draw[eDot] (v-C-1) to (v-C-2);
\end{tikzpicture}
};
\node[vCat, right=of C] (A) {
\begin{tikzpicture}[gDot, remember picture]
\node[vDot] (v-A-1) {};
\node[vDot, below=of v-A-1] (v-A-2) {};
\end{tikzpicture}
};
\node[vCat, right=of A] (B) {
\begin{tikzpicture}[gDot, remember picture]
\node[vDot] (v-B-1) {};
\node[vDot, below=of v-B-1] (v-B-2) {};
\draw[eDot] (v-B-1) to (v-B-2);
\end{tikzpicture}
};
\draw[eDotMorphism] (v-A-1) to (v-B-1);
\draw[eDotMorphism] (v-A-2) to (v-B-2);
\draw[eDotMorphism] (v-A-1) to (v-C-1);
\draw[eDotMorphism] (v-A-2) to (v-C-2);
\end{tikzpicture}
}
\qquad
\subcaptionbox{\label{fig:pushout:simple:p}}{
\begin{tikzpicture}[gCat]
\node[vCat] {
\begin{tikzpicture}[gDot, remember picture]
\node[vDot] (v-B-1) {};
\node[vDot, below=of v-B-1] (v-B-2) {};
\draw[eDot] (v-B-1) to (v-B-2);
\end{tikzpicture}
};
\end{tikzpicture}
}
\qquad
\subcaptionbox{\label{fig:pushout:simple:parallel}}{
\begin{tikzpicture}[gCat]
\node[vCat] {
\begin{tikzpicture}[gDot, remember picture]
\node[vDot] (v-B-1) {};
\node[vDot, below=of v-B-1] (v-B-2) {};
\draw[eDot] (v-B-1) to[bend right=45] (v-B-2);
\draw[eDot] (v-B-1) to[bend right=-45] (v-B-2);
\end{tikzpicture}
};
\end{tikzpicture}
}
\caption{
The pushout object of Fig.~\subref{fig:pushout:simple:span}, in the category of simple graphs is either not existing or is the graph depicted in Fig.~\subref{fig:pushout:simple:p}, where the two edges are merged.
For multigraphs the pushout object would be the graph depicted in Fig.~\subref{fig:pushout:simple:parallel}.
}
\label{fig:pushout:simple}
\end{figure}

The choice of disallowing parallel edges is motivated by the aim of modelling of chemistry, where bonds between atoms are single entities.
While a ``double bond'' consists of twice the amount of electrons than a ``single bond'', it does not in general behave as two single bonds.
However, when parallel edges are disallowed a special situation arises when constructing pushouts.
Consider the span in Fig.~\ref{fig:pushout:simple:span}.
If parallel edges are allowed, the pushout object is the one shown in Fig.~\ref{fig:pushout:simple:parallel}.
Without parallel edges we could identify the edges as shown in Fig.~\ref{fig:pushout:simple:p}.
This approach was used in for example \cite{pushoutSimpleGraphs}.
However, for chemistry this means that we must define how to add two bonds together, which is not meaningful.
We therefore simply define that no pushout object exists for the span.
A direct derivation with the Double Pushout approach thus additionally requires that the second pushout is defined.

The explicit use of multisets gives rise to a form of minimality of a
derivation.  If $\mset{g_a, g_b, g_b}\xRightarrow{p, m} \mset{h_c, h_d}$ is
a valid derivation, for some rule $p$ and match $m$, then the extended
derivation $\mset{g_a, g_b, g_b, q}\xRightarrow{p, m} \mset{h_c, h_d, q}$
is also valid, even though $q$ is not ``used''.  We therefore say that a
derivation $G\xRightarrow{p, m} H$ with the left-hand side $G = \mset{g_1,
  g_2, \dots, g_n}$ is \emph{proper} if and only if
\begin{align*}
  g_i\cap \img(m) \neq \emptyset, \forall 1\leq i\leq n
\end{align*}
That is, if all connected components of $G$ are hit by the match.
The algorithms in \modAbbr{} only enumerate proper derivations.

\section{Graphs and Rules}
\label{sec:graphsRules}
Graphs and rules are available as classes in the library.
A rule $(L\xleftarrow{l} K\xrightarrow{r} R)$ can be loaded from a description in GML \cite{gml} format.
As both $l$ and $r$ are monomorphisms the rule is represented without redundant information in GML
by three sets corresponding somewhat to the graph fragments $L\backslash K$, $K$, and $R\backslash K$
(see Fig.~\ref{fig:graphsRules} for details).

Graphs can similarly be loaded from GML descriptions,
and molecule graphs can also be loaded using the SMILES format \cite{smiles}
where most hydrogen atoms are implicitly specified.
A SMILES string is a pre-order recording of a depth-first traversal of the connected graph,
where back-edges are replaced with pairs of integers.

Both input methods result in objects which internally
stores the graph structure, where all labels are text strings.
Figure~\ref{fig:graphsRules} shows examples of graph and rule loading, using the
Python interface of the software.
\begin{figure}[tbp]
\centering
\begin{lstlisting}
formaldehyde = graphGML("formaldehyde.gml")
caffeine = smiles("Cn1cnc2c1c(=O)n(c(=O)n2C)C")
ketoEnol = ruleGMLString("""rule [
   left [
      edge [ source 1 target 4 label "-" ]
      edge [ source 1 target 2 label "-" ]
      edge [ source 2 target 3 label "=" ]
      node [ id 3 label "O" ]
      node [ id 4 label "H" ]
   ]
   context [
      node [ id 1 label "C" ]
      node [ id 2 label "C" ]
   ]
   right [
      edge [ source 1 target 2 label "=" ]
      edge [ source 2 target 3 label "-" ]
      node [ id 3 label "O-" ]
      node [ id 4 label "H+" ]
   ]
]""")
\end{lstlisting}
\caption[]{
Creation of two graph objects and a transformation rule object in the Python interface.
The (molecule) graph \lit{formaldehyde} is loaded from an external GML file,
while the (molecule) graph \lit{caffeine} is loaded from a SMILES string\cite{smiles},
often used in cheminformatics.
General labelled graphs can only be loaded from a GML description,
and all graphs are internally stored simply as labelled adjacency lists.
The DPO transformation rule \lit{ketoEnol} is loaded form an inline GML description.
When the GML sections \lit{left}, \lit{context}, and \lit{right} are considered sets,
they encode a rule $(L\leftarrow K\rightarrow R)$ with
$L = \text{\lit{left}}\cup \text{\lit{context}}$, $R = \text{\lit{right}}\cup \text{\lit{context}}$, and $K = \text{\lit{context}} \cup (\text{\lit{left}} \cap \text{\lit{right)}}$.
Vertices and edges that change label are thus specified in both \lit{left} and \lit{right}.
Note that in GML the endpoints of edges are described by \lit{source} and \lit{target},
but for undirected graphs these tags have no particular meaning and may be exchanged.
The graphs and rules are visualised in Fig.~\ref{fig:graphRulePrinting}.
}
\label{fig:graphsRules}
\end{figure}

Graphs have methods for counting both monomorphisms and isomorphisms, e.g.,
for substructure search and for finding duplicate graphs.  Counting the
number of carbonyl groups in a molecule \lit{mol} can be done simply as
\\\begin{minipage}{\linewidth}
\begin{lstlisting}
carbonyl = smiles("[C]=O")
count = carbonyl.monomorphism(mol, maxNumMatches=1337)
\end{lstlisting}
\end{minipage}
By default the \lit{monomorphism} method stops searching after the first
morphism is found; alternative matches can be retrieved by setting the
limit to a higher value.

Rule objects also have methods for counting monomorphisms and isomorphisms.
A rule morphism $m\colon p_1\rightarrow p_2$ on the rules $p_i =
(L_i\xleftarrow{l_i} K_i\xrightarrow{r_i} R_i), i = 1, 2$ is a 3-tuple of
graph morphisms $m_X\colon X_1\rightarrow X_2, X\in \{L, K, R\}$ such that
they commute with the morphisms in the rules.  Finding an isomorphism
between two rules can thus be used for detecting duplicate rules, while
finding a monomorphism $m\colon p_1\rightarrow p_2$ determines that $p_1$
is at least as general as $p_2$.

\section{Composition of Transformation Rules}
\label{sec:ruleComp}
In \cite{ruleComp,trace} the concept of rule composition is described,
where two rules $p_1 = (L_1\leftarrow K_1\rightarrow R_1), p_2 =
(L_2\leftarrow K_2\rightarrow R_2)$ are composed along a common subgraph
given by the span $R_1\leftarrow D\rightarrow L_2$.
Different types of rule composition can be defined by restricting the common subgraph and its
relation to the two rules. \modAbbr{} implements enumeration algorithms for
several special cases that are motived and defined in \cite{ruleComp,trace}.
The simplest case is to set $D$ as the empty graph, denoted by the operator $\rcParallel$, to create a composed rule that implements the parallel application of two rules.
In the most general case, denoted by $\rcCommon$, all common subgraphs of $R_1$ and $L_2$ are enumerated.
In a more restricted setting $R_1$ is a subgraph of $L_2$, denoted by $\rcFullRev$,
or, symmetrically, $L_2$ is a subgraph of $R_1$, denoted by $\rcFull$.
When the subgraph requirement is relaxed to only hold for a subset of the connected components of the graphs we denoted it by $\rcPartialRev$ and~$\rcPartial$.

The Python interface contains a mini-language for computing the result of rule
composition expressions with these operators.
The grammar for this language of expressions is shown in Fig.~\ref{fig:rc:grammar}.
\begin{figure}[tbp]
\centering
\scriptsize
\subcaptionbox{\label{fig:rc:grammar:grammar}}{
\settowidth\grammarindent{$\langle rcExp\rangle$\hspace{1em}}
\begin{minipage}{0.45\textwidth}
\begin{grammar}
<rcExp> :: \synt{rcExp} \synt{op} \synt{rcExp}
	\alt \lit{rcBind(} \synt{graphs} \lit{)}
	\alt \lit{rcUnbind(} \synt{graphs} \lit{)}
	\alt \lit{rcId(} \synt{graphs} \lit{)}
	\alt \synt{rules}
\end{grammar}
\end{minipage}
}
\hfill
\subcaptionbox{\label{fig:rc:grammar:op}}{
\begin{tabular}{@{}ll@{}}
\toprule
Math Operator	& Non-terminal \synt{op}			\\
\midrule
$\rcParallel$	& \lit{*rcParallel*}				\\
$\rcFull$		& \lit{*rcSuper(allowPartial=False)*}	\\
$\rcPartial$		& \lit{*rcSuper*}			\\
$\rcFullRev$	& \lit{*rcSub(allowPartial=False)*}	        \\
$\rcPartialRev$	& \lit{*rcSub*}					\\
$\rcCommon$	& \lit{*rcCommon*}				\\
\bottomrule
\end{tabular}
}
\caption[]{Grammar for rule composition expressions in the Python
  interface, where \synt{graphs} is a Python expression returning either a
  single graph or a collection of graphs.  Similarly is \synt{rules} a
  Python expression returning either a single rule or a collection of
  rules.  The pseudo-operators \synt{op} each correspond to a mathematical
  rule composition operator (see \cite{ruleComp,trace}).  The three
  functions \lit{rcBind}, \lit{rcUnbind}, and \lit{rcId} refers to the
  construction of the respective rules
	$(\emptyset\leftarrow \emptyset\rightarrow G)$,
	$(G\leftarrow \emptyset\rightarrow \emptyset)$, and
	$(G\leftarrow G\rightarrow G)$ from a graph $G$.
}
\label{fig:rc:grammar}
\end{figure}

Its implementation is realised using a series of global objects with
suitable overloading of the multiplication operator.  A rule composition
expression can be passed to an evaluator, which will carry out the
composition and discard duplicate results, as determined by checking
isomorphism between rules.  The result of each \synt{rcExp} is coerced into
a list of rules, and the operators consider all selections of rules from
their arguments.  That is, if \lit{P1} and \lit{P2} are two rule
composition expressions, whose evaluation results in two corresponding
lists of rules, $P_1$ and $P_2$.  Then, for example, the evaluation of
\lit{P1 *rcParallel* P2} results in the following list of rules:
\begin{align*}
    \bigcup_{p_1\in P_1}\bigcup_{p_2\in P_2}p_1\rcParallel p_2
\end{align*}
Each of these rules encodes the parallel application of a rule from $P_1$
and a rule from $P_2$.

In the following Python code, for example, we compute the rules corresponding to the bottom span
$(G\leftarrow D\rightarrow H)$ of a DPO diagram,
arising from applying the rule $p = (L\leftarrow K\rightarrow R)$ to the multiset of connected graphs
$G = \mset{g_1, g_2}$.
\\\begin{minipage}{\linewidth} 
\begin{lstlisting}
exp = rcId(g1) *rcParallel* rcId(g2) *rcSuper(allowPartial=False)* p
rc = rcEvaluator(ruleList)
res = rc.eval(exp)
\end{lstlisting}
\end{minipage}
Here, the rule composition evaluator is given a list \lit{ruleList} of
known rules that will be used for detecting isomorphic rules.  Larger rule
composition expressions, such as those found in \cite{trace}, can similarly
be directly written as Python code.

\section{Exploration of Graph Languages Using Strategies}
\label{sec:dgStrat}

A breadth-first enumeration of the language of a graph grammar is not
always desirable.  For example, in chemical systems there are often
constraints that can not be expressed easily in the underlying graph
transformation rules.
In~\cite{dgStrat} a strategy framework is introduced for the exploration of graph languages.
It is a domain specific programming language that, like the rule composition expressions,
is implemented in the Python interface, with the grammar shown in Fig.~\ref{fig:dg:grammar}.
\begin{figure}[tbp]
\centering
\settowidth\grammarindent{$\langle rcExp\rangle$\hspace{0.8em}}
\begin{grammar}
<strat> :: <strats> | <strat> `\verb!>>!' <strat> | <rule>
	\alt \lit{addSubset(} <graphs> \lit{)} | \lit{addUniverse(} <graphs> \lit{)}
	\alt \lit{filterSubset(} <filterPred> \lit{)} | \lit{filterUniverse(} <filterPred> \lit{)}
	\alt \lit{leftPredicate[} <derivationPred> \lit{](} <strat> \lit{)}
	\alt \lit{rightPredicate[} <derivationPred> \lit{](} <strat> \lit{)}
	\alt \lit{repeat} [ \lit{[} <int> \lit{]} ] \lit{(} <strat> \lit{)}
	\alt \lit{revive(} <strat> \lit{)}
\end{grammar}
\cprotect\caption[]{%
  Grammar for the domain specific language for guiding graph
  transformation, embedded in the Python interface of the software package.
  The non-terminal \synt{strats} must be a collection of strategies,
  that becomes a $\stratParallel$ strategy from \cite{dgStrat}.  The
  production \synt{strat} `\verb!>>!' \synt{strat} results in a sequence
  strategy.
}
\label{fig:dg:grammar}
\end{figure}
The language computes on sets of graphs.
Simplified, this means that each execution state is a set of connected graphs.
An \emph{addition strategy} adds further graphs to this state,
and a \emph{filter strategy} removes graphs from it.
A \emph{rule strategy} enumerates direct derivations based on the state,
subject to acceptance by filters introduced by the \emph{left-} and \emph{right-predicate strategies}.
Newly derived graphs are added to the state.
Strategies can be sequentially composed with the `\verb!>>!' operator,
which can be extended to $k$-fold composition with the \emph{repetition strategy}.
A \emph{parallel strategy} executes multiple strategies with the same input, and merges their output.
During the execution of a program the discovered direct derivations are recorded as an annotated directed multi-hypergraph, which for chemical systems is a reaction network.
For a full definition of the language see~\cite{dgStrat} or the \modAbbr{} documentation.

A strategy expression must, similarly to a rule composition expression, be
given to an evaluator which ensures that isomorphic graphs are represented
by the same C++/Python object.  After execution the evaluator contains the
generated derivation graph, which can be visualised or programmatically
used for subsequent analysis.

The strategy language can for example be used for the simple breadth-first
exploration of a grammar with a set of graphs \lit{startingGraphs} and a set of rules
\lit{ruleSet}, where exploration does not result in graphs above a certain size (42 vertices):
\\\begin{minipage}{\linewidth} 
\begin{lstlisting}
strat = (
	   addSubset(startingGraphs)
	>> rightPredicate[
		lambda derivation: all(g.numVertices <= 42 for g in derivation.right)
	](    repeat(ruleSet)    )
)
dg = dgRuleComp(startingGraphs, strat)
dg.calc()
\end{lstlisting}
\end{minipage}
The \lit{dg} object is the evaluator which afterwards contains the
derivation graph.  More examples can be found in \cite{dgStrat} and
\cite{hcn} where complex chemical behaviour is incorporated into
strategies.  An abstract example can also be found in \cite{dgStrat} where
the puzzle game Catalan \cite{catalan} is solved using exploration
strategies.

\section{Figure Generation}
\label{sec:figgen}

The software package includes elaborate functionality for automatically
visualising graph, rules, derivation graphs, and derivations.  The final
rendering of figures is done using the TikZ \cite{tikz} package for
\LaTeX{}, while the layouts for graphs are computed using Graphviz
\cite{graphviz}.  However, for molecule graphs it is possible to use the
cheminformatics library Open Babel \cite{obabel} for laying out molecules
and reaction patterns in a more chemically familiar manner.

Visualisation starts by calling a \lit{print} method on the object in
question.  This generates files with \LaTeX{} code and a graph description
in Graphviz format.  Special post-processing commands are additionally
inserted into another file.  Invoking the post-processor will then generate
coordinates and compile the final layout.  In addition, an aggregate
summary document is compiled that includes all figures for easy overview.
Fig.~\ref{fig:graphRulePrinting} shows an example, where the wrapper
script \lit{mod} provided by the package is used to automatically execute
both a Python script and subsequently the post-processor.
\begin{figure}[tbp]
\centering
\begin{lrbox}{\listingsBox}
\begin{minipage}{0.42\textwidth}
\begin{lstlisting}
p = GraphPrinter()
p.setMolDefault()
p.collapseHydrogens = False
formaldehyde.print(p)
p.edgesAsBonds = False
caffeine.print(p)
p.setReactionDefault()
ketoEnol.print(p)
\end{lstlisting}
\end{minipage}
\end{lrbox}
\subcaptionbox{
	Additional Python code to Fig.~\ref{fig:graphsRules},
	for generating figures.
}{%
	\usebox{\listingsBox}
}%
\hfill
\subcaptionbox{\label{fig:graphRulePrinting:graph}
	Automatically compiled figure of the two graphs loaded in Fig.~\ref{fig:graphsRules}.
}[0.42\textwidth]{%
	\raisebox{-0.5\height}{%
    		\graphGML[collapse hydrogens=false][scale=0.4]{data/formaldehyde.gml}
	}
	\hfill
	\raisebox{-0.5\height}{%
		\smiles[collapse hydrogens=false, edges as bonds=false][scale=0.4]{Cn1cnc2c1c(=O)n(c(=O)n2C)C}
	}
}

\subcaptionbox{\label{fig:graphRulePrinting:rule}
	Automatically compiled figure of the DPO rule loaded in Fig.~\ref{fig:graphsRules}.
}[\textwidth]{%
	{\scriptsize\ruleGML{data/ketoEnol.gml}{\dpoRule[scale=0.4]}}
}
\cprotect\caption[]{%
	Example of automatic visualisation of graphs and rules, using the post-processor.
	The Python code is an extension of the code from Fig.~\ref{fig:graphsRules},
	and can be executed using the provided \lit{mod} script that invokes both
	the Python interpreter \lit{python3} and the post-processor, \lit{mod\_post}.
	Edges with special labels are as default rendered in a special chemical manner,
	as illustrated with the left graph of Fig.~\subref{fig:graphRulePrinting:graph} (formaldehyde).
	In the right graph of Fig.~\subref{fig:graphRulePrinting:graph} (caffeine) the edge labels are shown
	explicitly. Both graphs uses chemical colouring.
	The colouring of the transformation rule, Fig.~\subref{fig:graphRulePrinting:rule},
	denote the differences between $L$, $K$, and $R$.
}
\label{fig:graphRulePrinting}
\end{figure}
The example also shows part of the functionality for chemical rendering
options, such as atom-specific colouring, charges rendered in superscript,
and collapsing of hydrogen vertices into their neighbours.

\begin{figure}[tbp]
\centering
\begin{lrbox}{\listingsBox}
\begin{minipage}{0.92\textwidth}
\begin{lstlisting}
p = DGPrinter()
p.pushVertexLabel(lambda g, dg: "|V| = %d" % g.numVertices)
p.pushVertexColour(lambda g, dg: "blue" if g.numVertices >= 16 else "")
dg.print(p)
\end{lstlisting}
\end{minipage}
\end{lrbox}
\subcaptionbox{
  Python code for customised visualisation of a derivation graph \lit{dg}.
}{\usebox{\listingsBox}}
\subcaptionbox{
	Example of automatically laid out and rendered derivation  graph with custom labelling and colour.
}{
  \includegraphics[scale=0.3]{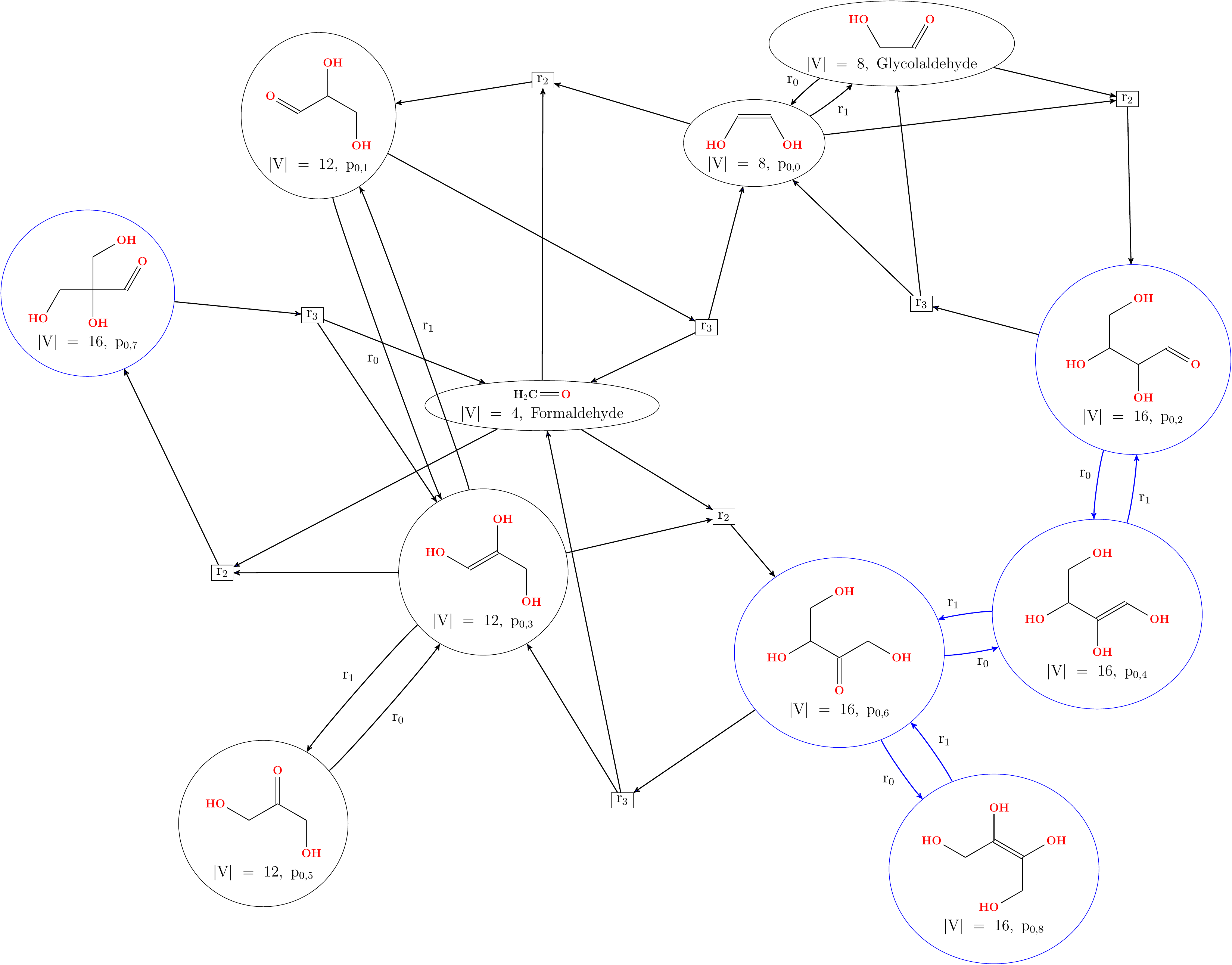}
}
\caption[]{%
  Example of derivation graph printing. Each vertex is as default labelled
  with the name of the graph it represents, and a figure of the graph is
  embedded.  Each hyperedge is as default labelled with the name of the
  rule used in the derivation the hyperedge represents.
  A general hyperedge is represented by a box, but for hyperedges with only 1 head and 1 tail
  the box is omitted, and a single labelled arc is rendered.
}
\label{fig:dgPrinting}
\end{figure}

Derivation graphs can also be visualised automatically, where each vertex
is depicted with a rendering of the graph it represents.  The overall
depiction can be customised to a high degree, e.g., by annotation or
colouring of vertices and hyperedges using user-defined callback functions.
Fig.~\ref{fig:dgPrinting} illustrates part of this functionality.

\begin{figure}[tbp]
\centering
\begin{lrbox}{\listingsBox}
\begin{minipage}{0.43\textwidth}
\begin{lstlisting}
for dRef in dg.derivations:
	dRef.print()
\end{lstlisting}
\end{minipage}
\end{lrbox}
\subcaptionbox{
  Python code for visualising all derivations in a derivation graph \lit{dg}.
}{\usebox{\listingsBox}}
\hfill
\subcaptionbox{An automatically generated Double Pushout diagram.}{
\scriptsize
\derivation[scale=0.32]
	{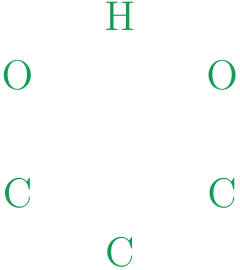}
	{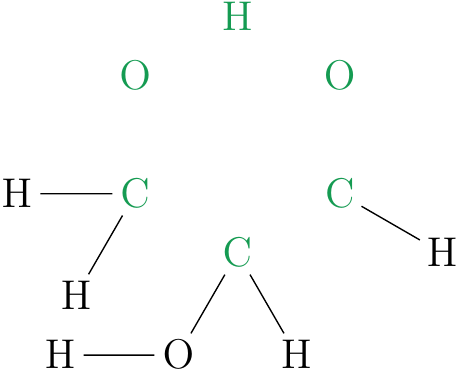}
}
\caption[]{
Example of visualisation of derivations.
Each derivation from a derivation graph can be printed, with the same
customisation options as for graphs and rules.  Additional colouring is
used to highlight the image of the rule into the lower span.  }
\label{fig:dpoDiagram}
\end{figure}

Individual derivations of a derivation graph can be visualised in form of
Double Pushout diagrams.  The rendering of these diagrams can be customised
similar to how rules and graph depictions can, e.g., to make the graphs
have a more chemical feel.  An example of derivation printing is
illustrated in Fig.~\ref{fig:dpoDiagram}.

\begin{figure}[tbp]
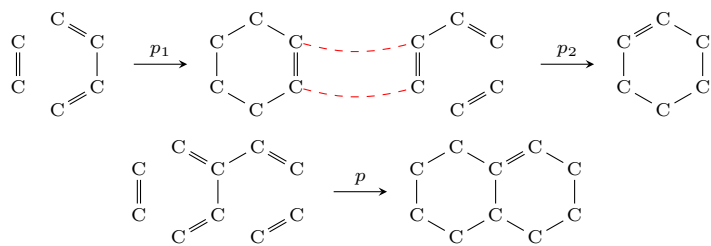

\centering
\renewcommand{\modInputPrefix}{figures/rcMatch}
\newcommand{\rcMatch}[1]{%
\begin{minipage}[t]{\textwidth}%
\scriptsize%
\renewcommand{\modGraphScale}{0.6}%
#1%
\end{minipage}%
}
\rcMatch{\centering
\begin{tikzpicture}[remember picture, scale=\modGraphScale, baseline={([yshift={-\ht\strutbox}]current bounding box)}]
\input{\modInputPrefix/out/144_r_93.coord.tex}
\node[modStyleGraphVertex, at=(v-coord-0)] (v-0) {C};
\node[modStyleGraphVertex, at=(v-coord-1)] (v-1) {C};
\node[modStyleGraphVertex, at=(v-coord-2)] (v-2) {C};
\node[modStyleGraphVertex, at=(v-coord-3)] (v-3) {C};
\node[modStyleGraphVertex, at=(v-coord-4)] (v-4) {C};
\node[modStyleGraphVertex, at=(v-coord-5)] (v-5) {C};
\modDrawDoubleBond{0}{1}{1}{1}{}
\modDrawSingleBond{1}{2}{1}{1}{}
\modDrawDoubleBond{2}{3}{1}{1}{}
\modDrawDoubleBond{4}{5}{1}{1}{}
\end{tikzpicture}
\begin{tikzpicture}[node distance=20pt,
				baseline={([yshift={-\ht\strutbox}]A.center)}]
				\node (A) {};
				\node (B) [right=of A] {};
				\draw[eMorphism] (A) to node[above] {$p_1$} (B);
				\end{tikzpicture}
\begin{tikzpicture}[remember picture, scale=\modGraphScale, baseline={([yshift={-\ht\strutbox}]current bounding box)}]
\input{\modInputPrefix/out/144_r_93.coord.tex}
\node[modStyleGraphVertex, at=(v-coord-0)] (v-0) {C};
\node[modStyleGraphVertex, at=(v-coord-1)] (v-1) {C};
\node[modStyleGraphVertex, at=(v-coord-2)] (v-2) {C};
\node[modStyleGraphVertex, at=(v-coord-3)] (v-3) {C};
\node[modStyleGraphVertex, at=(v-coord-4)] (v-4) {C};
\node[modStyleGraphVertex, at=(v-coord-5)] (v-5) {C};
\modDrawSingleBond{0}{1}{1}{1}{}
\modDrawDoubleBond{1}{2}{1}{1}{}
\modDrawSingleBond{2}{3}{1}{1}{}
\modDrawSingleBond{3}{4}{1}{1}{}
\modDrawSingleBond{4}{5}{1}{1}{}
\modDrawSingleBond{5}{0}{1}{1}{}
\end{tikzpicture}
\hspace{4em}
\begin{tikzpicture}[remember picture, scale=\modGraphScale, baseline={([yshift={-\ht\strutbox}]current bounding box)}]
\input{\modInputPrefix/out/146_r_94.coord.tex}
\node[modStyleGraphVertex, at=(v-coord-6)] (v-6) {C};
\node[modStyleGraphVertex, at=(v-coord-7)] (v-7) {C};
\node[modStyleGraphVertex, at=(v-coord-8)] (v-8) {C};
\node[modStyleGraphVertex, at=(v-coord-9)] (v-9) {C};
\node[modStyleGraphVertex, at=(v-coord-10)] (v-10) {C};
\node[modStyleGraphVertex, at=(v-coord-11)] (v-11) {C};
\modDrawDoubleBond{6}{7}{1}{1}{}
\modDrawSingleBond{7}{8}{1}{1}{}
\modDrawDoubleBond{8}{9}{1}{1}{}
\modDrawDoubleBond{10}{11}{1}{1}{}
\end{tikzpicture}
\begin{tikzpicture}[remember picture, overlay]
\path[modRCMatchEdge] (v-1) to[modRCMatchEdgeTo] (v-7);
\path[modRCMatchEdge] (v-2) to[modRCMatchEdgeTo] (v-6);
\end{tikzpicture}
\begin{tikzpicture}[node distance=20pt,
				baseline={([yshift={-\ht\strutbox}]A.center)}]
				\node (A) {};
				\node (B) [right=of A] {};
				\draw[eMorphism] (A) to node[above] {$p_2$} (B);
				\end{tikzpicture}
\begin{tikzpicture}[remember picture, scale=\modGraphScale, baseline={([yshift={-\ht\strutbox}]current bounding box)}]
\input{\modInputPrefix/out/146_r_94.coord.tex}
\node[modStyleGraphVertex, at=(v-coord-6)] (v-6) {C};
\node[modStyleGraphVertex, at=(v-coord-7)] (v-7) {C};
\node[modStyleGraphVertex, at=(v-coord-8)] (v-8) {C};
\node[modStyleGraphVertex, at=(v-coord-9)] (v-9) {C};
\node[modStyleGraphVertex, at=(v-coord-10)] (v-10) {C};
\node[modStyleGraphVertex, at=(v-coord-11)] (v-11) {C};
\modDrawSingleBond{6}{7}{1}{1}{}
\modDrawDoubleBond{7}{8}{1}{1}{}
\modDrawSingleBond{8}{9}{1}{1}{}
\modDrawSingleBond{9}{10}{1}{1}{}
\modDrawSingleBond{10}{11}{1}{1}{}
\modDrawSingleBond{11}{6}{1}{1}{}
\end{tikzpicture}
\\[1em]
\begin{tikzpicture}[remember picture, scale=\modGraphScale, baseline={([yshift={-\ht\strutbox}]current bounding box)}]
\input{\modInputPrefix/out/148_r_95.coord.tex}
\node[modStyleGraphVertex, at=(v-coord-0)] (v-0) {C};
\node[modStyleGraphVertex, at=(v-coord-1)] (v-1) {C};
\node[modStyleGraphVertex, at=(v-coord-2)] (v-2) {C};
\node[modStyleGraphVertex, at=(v-coord-3)] (v-3) {C};
\node[modStyleGraphVertex, at=(v-coord-4)] (v-4) {C};
\node[modStyleGraphVertex, at=(v-coord-5)] (v-5) {C};
\node[modStyleGraphVertex, at=(v-coord-6)] (v-6) {C};
\node[modStyleGraphVertex, at=(v-coord-7)] (v-7) {C};
\node[modStyleGraphVertex, at=(v-coord-8)] (v-8) {C};
\node[modStyleGraphVertex, at=(v-coord-9)] (v-9) {C};
\modDrawDoubleBond{0}{1}{1}{1}{}
\modDrawSingleBond{1}{2}{1}{1}{}
\modDrawDoubleBond{2}{3}{1}{1}{}
\modDrawDoubleBond{4}{5}{1}{1}{}
\modDrawSingleBond{1}{6}{1}{1}{}
\modDrawDoubleBond{6}{7}{1}{1}{}
\modDrawDoubleBond{8}{9}{1}{1}{}
\end{tikzpicture}
\begin{tikzpicture}[node distance=20pt,
				baseline={([yshift={-\ht\strutbox}]A.center)}]
				\node (A) {};
				\node (B) [right=of A] {};
				\draw[eMorphism] (A) to node[above] {$p$} (B);
				\end{tikzpicture}
\begin{tikzpicture}[remember picture, scale=\modGraphScale, baseline={([yshift={-\ht\strutbox}]current bounding box)}]
\input{\modInputPrefix/out/148_r_95.coord.tex}
\node[modStyleGraphVertex, at=(v-coord-0)] (v-0) {C};
\node[modStyleGraphVertex, at=(v-coord-1)] (v-1) {C};
\node[modStyleGraphVertex, at=(v-coord-2)] (v-2) {C};
\node[modStyleGraphVertex, at=(v-coord-3)] (v-3) {C};
\node[modStyleGraphVertex, at=(v-coord-4)] (v-4) {C};
\node[modStyleGraphVertex, at=(v-coord-5)] (v-5) {C};
\node[modStyleGraphVertex, at=(v-coord-6)] (v-6) {C};
\node[modStyleGraphVertex, at=(v-coord-7)] (v-7) {C};
\node[modStyleGraphVertex, at=(v-coord-8)] (v-8) {C};
\node[modStyleGraphVertex, at=(v-coord-9)] (v-9) {C};
\modDrawSingleBond{0}{1}{1}{1}{}
\modDrawSingleBond{1}{2}{1}{1}{}
\modDrawSingleBond{2}{3}{1}{1}{}
\modDrawSingleBond{3}{4}{1}{1}{}
\modDrawSingleBond{4}{5}{1}{1}{}
\modDrawSingleBond{5}{0}{1}{1}{}
\modDrawDoubleBond{1}{6}{1}{1}{}
\modDrawSingleBond{6}{7}{1}{1}{}
\modDrawSingleBond{7}{8}{1}{1}{}
\modDrawSingleBond{8}{9}{1}{1}{}
\modDrawSingleBond{9}{2}{1}{1}{}
\end{tikzpicture}}
\caption[]{ 
  Visualisation of the composition of two rules $p_i = (L_i\leftarrow
  K_i\rightarrow R_i, i = 1,2$, along the a common subgraph of $R_1$ and
  $L_2$, indicated by the dashed red lines.  Only the left and right graphs
  of both rules, and the resulting rule, are shown. The rendering can be
  customised in the same manner as the rendering for graphs and rules can.
}
\label{fig:rc:match}
\end{figure}

Composition of transformation rules is a core operation in the software,
and for better understanding the operation we provide a mechanism for
visualising individual compositions.  An example of such a visualisation is
shown in Fig.~\ref{fig:rc:match}, where only the left and right graphs of
two argument rules and the result rule are shown.
The composition relation is shown as red dashed lines between the left
graph of the first rule and the right graph of the second rule.

\subsubsection{Including Figures in \LaTeX{} Documents}
To make it easier to use illustrations of graphs and rules we have included
a \LaTeX{} package in the software.  It provides macros for automatically
generating Python scripts that subsequently generate figures and \LaTeX{}
code for inclusion into the original document.  For example, the depictions
in Fig.~\ref{fig:graphRulePrinting} are inserted with the following code.
\\\begin{minipage}{\linewidth}
\begin{lstlisting}[language=modtex]
\graphGML[collapse hydrogens=false][scale=0.4]{formaldehyde.gml}
\smiles[collapse hydrogens=false, edges as bonds=false][scale=0.4]
	{Cn1cnc2c1c(=O)n(c(=O)n2C)C}
\ruleGML{ketoEnol.gml}{\dpoRule[scale=0.4]}
\end{lstlisting}
\end{minipage}
Each \lit{\textbackslash graphGML} and \lit{\textbackslash smiles} macro
expands into an \lit{\textbackslash includegraphics} for a specific PDF
file, and a Python script is generated which can be executed to compile the
needed files.  The \lit{\textbackslash ruleGML} macro expands into
\\\begin{minipage}{\linewidth}
\begin{lstlisting}[language=modtex]
\dpoRule[scale=0.4]{fileL.pdf}{fileK.pdf}{fileR.pdf}
\end{lstlisting}
\end{minipage}
where the three PDF files depict the left side, context, and right side of
the rule.  The \lit{\textbackslash dpoRule} macro then expands into the
final rule diagram with the PDF files included.

\section{Summary}
\modName{} is a comprehensive software package for DPO graph transformation
on multisets of undirected, labelled graphs. It can be used for generic,
abstract graph models.  By providing many features for handling chemical
data it is particularly well-suited for modelling generative chemical
systems. The package includes an elaborate system for automatically
producing high-quality visualisations of graphs, rules, and DPO diagrams of
direct derivations.  

The first public version of \modAbbr{} described here is intended as the
foundation for a larger integrated package for graph-based cheminformatics.
Future versions will for example also include functionalities for pathway
analysis in reaction networks produced by the generative transformation methods
described here. The graph transformation system, on the other hand, will be
extended to cover more complicated chemical properties such as radicals,
charges, and stereochemistry. 

\section{Acknowledgements}
This work is supported by the Danish Council for Independent Research, Natural Sciences, the COST Action
CM1304 ``Emergence and Evolution of Complex Chemical Systems'', and
the ELSI Origins Network (EON), which is
supported by a grant from the John Templeton Foundation.
The opinions expressed in this publication are those of
the authors and do not necessarily reflect the views of the John Templeton
Foundation.

\bibliographystyle{plain}
\bibliography{paper}

\begin{thebibliography}{10}

\bibitem{hcn}
Jakob~L. Andersen, Tommy Andersen, Christoph Flamm, Martin~M. Hanczyc, Daniel
  Merkle, and Peter~F. Stadler.
\newblock Navigating the chemical space of hcn polymerization and hydrolysis:
  Guiding graph grammars by mass spectrometry data.
\newblock {\em Entropy}, 15(10):4066--4083, 2013.

\bibitem{ruleComp}
Jakob~L. Andersen, Christoph Flamm, Daniel Merkle, and Peter~F. Stadler.
\newblock Inferring chemical reaction patterns using rule composition in graph
  grammars.
\newblock {\em Journal of Systems Chemistry}, 4(1):4, 2013.

\bibitem{trace}
Jakob~L. Andersen, Christoph Flamm, Daniel Merkle, and Peter~F. Stadler.
\newblock 50 {S}hades of rule composition: From chemical reactions to higher
  levels of abstraction.
\newblock In Fran{\c{c}}ois Fages and Carla Piazza, editors, {\em Formal
  Methods in Macro-Biology}, volume 8738 of {\em Lecture Notes in Computer
  Science}, pages 117--135, Berlin, 2014. Springer International Publishing.

\bibitem{dgStrat}
Jakob~L. Andersen, Christoph Flamm, Daniel Merkle, and Peter~F. Stadler.
\newblock Generic strategies for chemical space exploration.
\newblock {\em International Journal of Computational Biology and Drug Design},
  7(2/3):225 -- 258, 2014.
\newblock {TR}: \url{http://arxiv.org/abs/1302.4006}.

\bibitem{Andrei11}
O.~Andrei, M.~Fern\'andez, H.~Kirchner, G.~Melan\c{c}on, O.~Namet, and
  B.~Pinaud.
\newblock {PORGY}: Strategy driven interactive transformation of graphs.
\newblock In {\em In Proceedings of the 6\textsuperscript{th} International
  Workshop on Computing with Terms and Graphs (TERMGRAPH 2011)}, volume~48 of
  {\em Electronic Proceedings in Theoretical Computer Science}, pages 54--68,
  2011.

\bibitem{benko}
Gil Benk{\"o}, Christoph Flamm, and Peter~F. Stadler.
\newblock A graph-based toy model of chemistry.
\newblock {\em Journal of Chemical Information and Computer Sciences},
  43(4):1085--1093, 2003.

\bibitem{pushoutSimpleGraphs}
Benjamin Braatz, Ulrike Golas, and Thomas Soboll.
\newblock How to delete categorically --- two pushout complement constructions.
\newblock {\em Journal of Symbolic Computation}, 46(3):246--271, 2011.
\newblock Applied and Computational Category Theory.

\bibitem{vf2:2}
L.P. Cordella, P.~Foggia, C.~Sansone, and M.~Vento.
\newblock A (sub) graph isomorphism algorithm for matching large graphs.
\newblock {\em IEEE Transactions on Pattern Analysis and Machine Intelligence},
  26(10):1367, 2004.

\bibitem{vf2:1}
Luigi~Pietro Cordella, Pasquale Foggia, Carlo Sansone, and Mario Vento.
\newblock An improved algorithm for matching large graphs.
\newblock In {\em Proc. of the 3\textsuperscript{rd} IAPR-TC15 Workshop on
  Graph-based Representations in Pattern Recognition}, pages 149--159, 2001.

\bibitem{handbook1}
A.~Corradini, U.~Montanari, F.~Rossi, H.~Ehrig, R.~Heckel, and M.~L{\"{o}}we.
\newblock {Algebraic Approaches to Graph Transformation -- Part I: Basic
  Concepts and Double Pushout Approach}.
\newblock In Grzegorz Rozenberg, editor, {\em Handbook of Graph Grammars and
  Computing by Graph Transformation}, chapter~3, pages 163--245. World
  Scientific, 1997.

\bibitem{Ehrig:2006}
Karsten Ehrig, Reiko Heckel, and Georgios Lajios.
\newblock Molecular analysis of metabolic pathway with graph transformation.
\newblock In Andrea Corradini, Hartmut Ehrig, Ugo Montanari, Leila Ribeiro, and
  Grzegorz Rozenberg, editors, {\em Graph Transformations: Third International
  Conference, ICGT 2006 Natal, Rio Grande do Norte, Brazil, September 17-23,
  2006 Proceedings}, pages 107--121. Springer, Berlin, Heidelberg, 2006.

\bibitem{Kirchner11}
M.~Fern{\'a}ndez, H.~Kirchner, and O.~Namet.
\newblock A strategy language for graph rewriting.
\newblock In {\em Proceedings of the 21\textsuperscript{st} International
  Symposium on Logic-Based Program Synthesis and Transformation (LOPSTR 2011)},
  volume 7225 of {\em Lecture Notes in Computer Science}, pages 173--188, 2012.

\bibitem{Flamm:10b}
Christoph Flamm, Alexander Ullrich, Heinz Ekker, Martin Mann, Daniel
  H{\"o}gerl, Markus Rohrschneider, Sebastian Sauer, Gerik Scheuermann,
  Konstantin Klemm, Ivo~L. Hofacker, and Peter~F. Stadler.
\newblock Evolution of metabolic networks: A computational framework.
\newblock {\em Journal of Systems Chemistry}, 1(4):4, 2010.

\bibitem{graphviz}
Emden~R. Gansner and Stephen~C. North.
\newblock An open graph visualization system and its applications to software
  engineering.
\newblock {\em SOFTWARE - PRACTICE AND EXPERIENCE}, 30(11):1203--1233, 2000.

\bibitem{gml}
M.~Himsolt.
\newblock {GML}: A portable graph file format.

\bibitem{catalan}
increpare games.
\newblock Catalan.
\newblock \url{http://www.increpare.com/2011/01/catalan/}, 2011.

\bibitem{Kreowski2011}
Hans-J{\"o}rg Kreowski and Sabine Kuske.
\newblock Graph multiset transformation: a new framework for massively parallel
  computation inspired by dna computing.
\newblock {\em Natural Computing}, 10(2):961--986, 2011.

\bibitem{ggl}
Martin Mann, Heinz Ekker, and Christoph Flamm.
\newblock The graph grammar library - a generic framework for chemical graph
  rewrite systems.
\newblock In Keith Duddy and Gerti Kappel, editors, {\em Theory and Practice of
  Model Transformations}, volume 7909 of {\em Lecture Notes in Computer
  Science}, pages 52--53. Springer Berlin Heidelberg, 2013.

\bibitem{obabel}
Noel~M O'Boyle, Michael Banck, Craig~A James, Chris Morley, Tim Vandermeersch,
  and Geoffrey~R Hutchison.
\newblock {Open Babel}: An open chemical toolbox.
\newblock {\em Journal of Cheminformatics}, 3(33), 2011.

\bibitem{Rossello:2004}
Francesc Rossell{\'o} and Gabriel Valiente.
\newblock Analysis of metabolic pathways by graph transformation.
\newblock In Hartmut Ehrig, Gregor Engels, Francesco Parisi-Presicce, and
  Grzegorz Rozenberg, editors, {\em Graph Transformations: Second International
  Conference, ICGT 2004, Rome, Italy, September 28--October 1, 2004.
  Proceedings}, pages 70--82, Berlin, Heidelberg, 2004. Springer.

\bibitem{Rossello:2005b}
Francesc Rossell{\'o} and Gabriel Valiente.
\newblock Chemical graphs, chemical reaction graphs, and chemical graph
  transformation.
\newblock {\em Electronic Notes in Theoretical Computer Science}, 127(1):157 --
  166, 2005.
\newblock Proceedings of the International Workshop on Graph-Based Tools
  (GraBaTs 2004)Graph-Based Tools 2004.

\bibitem{bgl}
Jeremy~G Siek, Lie-Quan Lee, and Andrew Lumsdaine.
\newblock {\em Boost Graph Library: The User Guide and Reference Manual}.
\newblock Pearson Education, 2001.
\newblock \url{http://www.boost.org/libs/graph/}.

\bibitem{sylvester}
J.~J. Sylvester.
\newblock On an application of the new atomic theory to the graphical
  representation of the invari- ants and covariants of binary quantics, with
  three appendices.
\newblock {\em American Journal of Mathematics}, 1(1):64--128, 1878.

\bibitem{Taentzer:2004}
Gabriele Taentzer.
\newblock {AGG}: A graph transformation environment for modeling and validation
  of software.
\newblock In John~L. Pfaltz, Manfred Nagl, and Boris B{\"o}hlen, editors, {\em
  Applications of Graph Transformations with Industrial Relevance: Second
  International Workshop, AGTIVE 2003, Charlottesville, VA, USA, September 27 -
  October 1, 2003, Revised Selected and Invited Papers}, pages 446--453,
  Berlin, Heidelberg, 2004. Springer.

\bibitem{tikz}
Till Tantau.
\newblock {\em The TikZ and PGF Packages}, 2013.

\bibitem{smiles}
D.~Weininger.
\newblock {SMILES, a chemical language and information system. 1. Introduction
  to methodology and encoding rules}.
\newblock {\em Journal of Chemical Information and Computer Sciences},
  28(1):31--36, 1988.

\bibitem{Yadav:2004}
ManeeshK. Yadav, BrianP. Kelley, and StevenM. Silverman.
\newblock The potential of a chemical graph transformation system.
\newblock In Hartmut Ehrig, Gregor Engels, Francesco Parisi-Presicce, and
  Grzegorz Rozenberg, editors, {\em Graph Transformations}, volume 3256 of {\em
  Lecture Notes in Computer Science}, pages 83--95. Springer Berlin Heidelberg,
  2004.

\end{thebibliography}

\appendix
\section{Examples}
The following is a short list of examples that show how \modName{} can be used via the Python interface.
They are all available as modifiable script in the live version of the software,
accessible at \url{http://mod.imada.sdu.dk/playground.html}.


\subsection{Graph Loading}
Molecules are encoded as labelled graphs.
They can be loaded from SMILES strings, and in general any graph can be loaded
from a GML specification, or from the SMILES-like format GraphDFS.
\lstset{basicstyle=\ttfamily\tiny}
\begin{lstlisting}
# Load a graph from a SMILES string (only for molecule graphs):
ethanol1 = smiles("CCO", name="Ethanol1")
# Load a graph from a SMILES-like format, called "GraphDFS", but for general graphs:
ethanol2 = graphDFS("[C]([H])([H])([H])[C]([H])([H])[O][H]", name="Ethanol2")
# The GraphDFS format also supports implicit hydrogens:
ethanol3 = graphDFS("CCO", name="Ethanol3")
# The basic graph format is GML:
ethanol4 = graphGMLString("""graph [
	node [ id 0 label "C" ]   node [ id 1 label "C" ]   node [ id 2 label "O" ]
	node [ id 3 label "H" ]   node [ id 4 label "H" ]   node [ id 5 label "H" ]
	node [ id 6 label "H" ]   node [ id 7 label "H" ]   node [ id 8 label "H" ]
	edge [ source 1 target 0 label "-" ]   edge [ source 2 target 1 label "-" ]
	edge [ source 3 target 0 label "-" ]   edge [ source 4 target 0 label "-" ]
	edge [ source 5 target 0 label "-" ]   edge [ source 6 target 1 label "-" ]
	edge [ source 7 target 1 label "-" ]   edge [ source 8 target 2 label "-" ]
]""", name="Ethanol4")
# They really are all loading the same graph into different objects:
assert ethanol1.isomorphism(ethanol2) == 1
assert ethanol1.isomorphism(ethanol3) == 1
assert ethanol1.isomorphism(ethanol4) == 1
# and they can be visualised:
ethanol1.print()
# All loaded graphs are added to a list 'inputGraphs':
for g in inputGraphs: g.print()
\end{lstlisting}

\subsection{Printing Graphs/Molecules}
The visualisation of graphs can be "prettified" using special printing options.
The changes can make the graphs look like normal molecule visualisations.
\lstset{basicstyle=\ttfamily\tiny}
\begin{lstlisting}
# Our test graph, representing the molecule caffeine:
g = smiles('Cn1cnc2c1c(=O)n(c(=O)n2C)C')
# ;ake an object to hold our settings:
p = GraphPrinter()
# First try visualising without any prettifications:
p.disableAll()
g.print(p)
# Now make chemical edges look like bonds, and put colour on atoms.
# Also put the "charge" part of vertex labels in superscript:
p.edgesAsBonds = True
p.raiseCharges=True
p.withColour = True
g.print(p)
# We can also "collapse" normal hydrogen atoms into the neighbours,
# and just show a count:
p.collapseHydrogens = True
g.print(p)
# And finally we can make "internal" carbon atoms simple lines:
p.simpleCarbons = True
g.print(p)
# There are also options for adding indices to the vertices,
# and modify the rendering of labels and edges:
p2 = GraphPrinter()
p2.disableAll()
p2.withTexttt = True
p2.thick = True
p2.withIndex = True
# We can actually print two different versions at the same time:
g.print(p2, p)
\end{lstlisting}

\subsection{Graph Interface}
Graph objects have a full interface to access individual vertices and edges.
The labels of vertices and edges can be accessed both in their raw string form,
and as their chemical counterpart (if they have one).
\lstset{basicstyle=\ttfamily\tiny}
\begin{lstlisting}
g = graphDFS("[R]{x}C([O-])CC=O")
print("|V| =", g.numVertices)
print("|E| =", g.numEdges)
for v in g.vertices:
	print("v%d: label='%s'" % (v.id, v.stringLabel), end="")
	print("\tas molecule: atomId=%d, charge=%d" % (v.atomId, v.charge), end="")
	print("\tis oxygen?", v.atomId == AtomIds.Oxygen)
	print("\td(v) =", v.degree)
	for e in v.incidentEdges: print("\tneighbour:", e.target.id)
for e in g.edges:
	print("(v%d, v%d): label='%s'" % (e.source.id, e.target.id, e.stringLabel), end="")
	try:
		bt = str(e.bondType)
	except LogicError:
		bt = "Invalid"
	print("\tas molecule: bondType=%s" % bt, end="")
	print("\tis double bond?", e.bondType == BondType.Double)
\end{lstlisting}

\subsection{Graph Morphisms}
Graph objects have methods for finding morphisms with the VF2 algorithms
for isomorphism and monomorphism.
We can therefore easily detect isomorphic graphs, count automorphisms,
and search for substructures.
\lstset{basicstyle=\ttfamily\tiny}
\begin{lstlisting}
mol1 = smiles("CC(C)CO")
mol2 = smiles("C(CC)CO")
# Check if there is just one isomorphism between the graphs:
isomorphic = mol1.isomorphism(mol2) == 1
print("Isomorphic?", isomorphic)
# Find the number of automorphisms in the graph,
# by explicitly enumerating all of them:
numAutomorphisms = mol1.isomorphism(mol1, maxNumMatches=1337)
print("|Aut(G)| =", numAutomorphisms)
# Let's count the number of methyl groups:
methyl = smiles("[CH3]")
# The symmetry of the group it self should not be counted,
# so find the size of the automorphism group of methyl.
numAutMethyl = methyl.isomorphism(methyl, maxNumMatches=1337)
print("|Aut(methyl)|", numAutMethyl)
# Now find the number of methyl matches,
numMono = methyl.monomorphism(mol1, maxNumMatches=1337)
print("#monomorphisms =", numMono)
# and divide by the symmetries of methyl.
print("#methyl groups =", numMono / numAutMethyl)
\end{lstlisting}

\subsection{Rule Loading}
Rules must be specified in GML format.
\lstset{basicstyle=\ttfamily\tiny}
\begin{lstlisting}
# A rule (L <- K -> R) is specified by three graph fragments:
# left, context, and right
destroyVertex = ruleGMLString('rule [   left    [   node [ id 1 label "A" ]   ]   ]')
createVertex = ruleGMLString( 'rule [   right   [   node [ id 1 label "A" ]   ]   ]')
identity = ruleGMLString(     'rule [   context [   node [ id 1 label "A" ]   ]   ]')
# A vertex/edge can change label:
labelChange = ruleGMLString("""rule [
	left    [   node [ id 1 label "A" ]   edge [ source 1 target 2 label "A" ]   ]
	# GML can have Python-style line comments too
	context [   node [ id 2 label "Q" ]                                          ]
	right   [   node [ id 1 label "B" ]   edge [ source 1 target 2 label "B" ]   ]
]""")
# A chemical rule should probably not destroy and create vertices:
ketoEnol = ruleGMLString("""rule [
	left [
		edge [ source 1 target 4 label "-" ]   edge [ source 1 target 2 label "-" ]
		edge [ source 2 target 3 label "=" ]
		node [ id 3 label "O" ]		node [ id 4 label "H" ]
	]   
	context [
		node [ id 1 label "C" ]   node [ id 2 label "C" ]
	]   
	right [
		edge [ source 1 target 2 label "=" ]   edge [ source 2 target 3 label "-" ]
		node [ id 3 label "O-" ]		node [ id 4 label "H+" ]
	]   
]""")
# Rules can be printed, but label changing edges are not visualised in K:
ketoEnol.print()
# Add with custom options, like graphs:
p1 = GraphPrinter()
p2 = GraphPrinter()
p1.disableAll()
p1.withTexttt = True
p1.withIndex = True
p2.setReactionDefault()
for p in inputRules: p.print(p1, p2)
# Be careful with printing options and non-existing implicit hydrogens:
p1.disableAll()
p1.edgesAsBonds = True
p2.setReactionDefault()
p2.simpleCarbons = True # !!
ketoEnol.print(p1, p2)
\end{lstlisting}

\subsection{Rule Morphisms}
Rule objects, like graph objects, have methods for finding morphisms with
the VF2 algorithms for isomorphism and monomorphism.
We can therefore easily detect isomorphic rules, 
and decide if one rule is at least as specific/general as another.
\lstset{basicstyle=\ttfamily\tiny}
\begin{lstlisting}
# A rule with no extra context:
small = ruleGMLString("""rule [         ruleID "Small"
	left  [   node [ id 1 label "H" ]    node [ id 2 label "O" ]   edge [ source 1 target 2 label "-" ]   ]
	right [   node [ id 1 label "H+" ]   node [ id 2 label "O-" ]                                         ]
]""")
# The same rule, with a bit of context:
large = ruleGMLString("""rule [           ruleID "Large"
	left    [   node [ id 1 label "H" ]    node [ id 2 label "O" ]   edge [ source 1 target 2 label "-" ]   ]
	context [   node [ id 3 label "C" ]    edge [ source 2 target 3 label "-" ]                             ]
	right   [   node [ id 1 label "H+" ]   node [ id 2 label "O-" ]
	]
]""")
isomorphic = small.isomorphism(large) == 1
print("Isomorphic?", isomorphic)
atLeastAsGeneral = small.monomorphism(large) == 1
print("At least as general?", atLeastAsGeneral)
\end{lstlisting}

\subsection{Formose Grammar}
The graph grammar modelling the formose chemistry.
\lstset{basicstyle=\ttfamily\tiny}
\begin{lstlisting}
formaldehyde = smiles("C=O", name="Formaldehyde")
glycolaldehyde = smiles( "OCC=O", name="Glycolaldehyde")
ketoEnolGML = """rule [   ruleID "Keto-enol isomerization" 
	left [      edge [ source 1 target 4 label "-" ]   edge [ source 1 target 2 label "-" ]
	            edge [ source 2 target 3 label "=" ]                                      ]
	context [   node [ id 1 label "C" ]   node [ id 2 label "C" ]
	            node [ id 3 label "O" ]   node [ id 4 label "H" ]                         ]
	right [     edge [ source 1 target 2 label "=" ]   edge [ source 2 target 3 label "-" ]
	            edge [ source 3 target 4 label "-" ]                                      ]   
]"""
ketoEnol_F = ruleGMLString(ketoEnolGML)
ketoEnol_B = ruleGMLString(ketoEnolGML, invert=True)
aldolAddGML = """rule [   ruleID "Aldol Addition"
	left [      edge [ source 1 target 2 label "=" ]   edge [ source 2 target 3 label "-" ]
	            edge [ source 3 target 4 label "-" ]   edge [ source 5 target 6 label "=" ]   ]
	context [   node [ id 1 label "C" ]   node [ id 2 label "C" ]   node [ id 3 label "O" ]
	            node [ id 4 label "H" ]   node [ id 5 label "O" ]   node [ id 6 label "C" ]   ]
	right [     edge [ source 1 target 2 label "-" ]   edge [ source 2 target 3 label "=" ]
	            edge [ source 5 target 6 label "-" ]
	            edge [ source 4 target 5 label "-" ]   edge [ source 6 target 1 label "-" ]   ]
]"""
aldolAdd_F = ruleGMLString(aldolAddGML)
aldolAdd_B = ruleGMLString(aldolAddGML, invert=True)
\end{lstlisting}

\subsection{Rule Composition 1 --- Unary Operators}
Special rules can be constructed from graphs.
\lstset{basicstyle=\ttfamily\tiny}
\begin{lstlisting}
glycolaldehyde.print()
# A graph G can be used to construct special rules:
# (\emptyset <- \emptyset -> G)
bindExp = rcBind(glycolaldehyde)
# (G <- \emptyset -> \emptyset)
unbindExp = rcUnbind(glycolaldehyde)
# (G <- G -> G)
idExp = rcId(glycolaldehyde)
# These are really rule composition expressions that have to be evaluated:
rc = rcEvaluator(inputRules)
# Each expression results in a lists of rules:
bindRules = rc.eval(bindExp)
unbindRules = rc.eval(unbindExp)
idRules = rc.eval(idExp)
postSection("Bind Rules")
for p in bindRules: p.print()
postSection("Unbind Rules")
for p in unbindRules: p.print()
postSection("Id Rules")
for p in idRules: p.print()
\end{lstlisting}

\subsection{Rule Composition 2 --- Parallel Composition}
A pair of rules can be merged to a new rule implementing the parallel transformation.
\lstset{basicstyle=\ttfamily\tiny}
\begin{lstlisting}
rc = rcEvaluator(inputRules)
# The special global object 'rcParallel' is used to make a pseudo-operator:
exp = rcId(formaldehyde) *rcParallel*  rcUnbind(glycolaldehyde)
rules = rc.eval(exp)
for p in rules: p.print()
\end{lstlisting}

\subsection{Rule Composition 3 --- Supergraph Composition}
A pair of rules can (maybe) be composed using a sueprgraph relation.
\lstset{basicstyle=\ttfamily\tiny}
\begin{lstlisting}
rc = rcEvaluator(inputRules)
exp = rcId(formaldehyde) *rcParallel*  rcId(glycolaldehyde)
exp = exp *rcSuper* ketoEnol_F
rules = rc.eval(exp)
for p in rules: p.print()
\end{lstlisting}

\subsection{Rule Composition 4 --- Overall Formose Reaction}
A complete pathway can be composed to obtain the overall rules.
\lstset{basicstyle=\ttfamily\tiny}
\begin{lstlisting}
rc = rcEvaluator(inputRules)
exp = (   rcId(glycolaldehyde) *rcSuper* ketoEnol_F
          *rcParallel* rcId(formaldehyde)
          *rcSuper(allowPartial=False)* aldolAdd_F
          *rcSuper* ketoEnol_F
          *rcParallel* rcId(formaldehyde)
          *rcSuper(allowPartial=False)* aldolAdd_F
          *rcSuper* ketoEnol_F
          *rcSuper* ketoEnol_B
          *rcSuper* aldolAdd_B
          *rcSuper* ketoEnol_B
          *rcSuper(allowPartial=False)*
          (rcId(glycolaldehyde) *rcParallel* rcId(glycolaldehyde))   )
rules = rc.eval(exp)
for p in rules: p.print()
\end{lstlisting}

\subsection{Reaction Networks 1 --- Rule Application}
Transformation rules (reaction patterns) can be applied to graphs (molecules) to create
new graphs (molecules). The transformations (reactions) implicitly form a
directed (multi-)hypergraph (chemical reaction network).
\lstset{basicstyle=\ttfamily\tiny}
\begin{lstlisting}
# Reaction networks are expaned using a strategy:
strat = (   # A molecule can be active or passive during evaluation.
            addUniverse(formaldehyde) # passive
            >> addSubset(glycolaldehyde) # active
            # Aach reaction must have a least 1 active educt.
            >> inputRules   )
# We call a reaction network a 'derivation graph'.
dg = dgRuleComp(inputGraphs, strat)
dg.calc()
# They can also be visualised.
dg.print()
\end{lstlisting}

\subsection{Reaction Networks 2 --- Repetition}
A sub-strategy can be repeated.
\lstset{basicstyle=\ttfamily\tiny}
\begin{lstlisting}
strat = (   addUniverse(formaldehyde)
            >> addSubset(glycolaldehyde)
            # Iterate the rule application 4 times.
            >> repeat[4](inputRules)   )
dg = dgRuleComp(inputGraphs, strat)
dg.calc()
dg.print()
\end{lstlisting}

\subsection{Reaction Networks 3 --- Application Constraints}
We may want to impose constraints on which reactions are accepted.
E.g., in formose the molecules should not have too many carbon atoms.
\lstset{basicstyle=\ttfamily\tiny}
\begin{lstlisting}
strat = (   addUniverse(formaldehyde)
            >> addSubset(glycolaldehyde)
            # Constrain the reactions:
            # No molecules with more than 20 atom can be created.
            >> rightPredicate[lambda derivation: all(g.numVertices <= 20 for g in derivation.right)](
               # Iterate until nothing new is found.
               repeat(inputRules)
            )   )
dg = dgRuleComp(inputGraphs, strat)
dg.calc()
dg.print()
\end{lstlisting}

\subsection{Advanced Printing}
Reaction networks can become large, and often it is necessary to hide parts of the network,
or in general change the appearance.
\lstset{basicstyle=\ttfamily\tiny}
\begin{lstlisting}
# Create a printer with default options:
p = DGPrinter()
# Hide "large" molecules: those with > 4 Cs:
p.pushVertexVisible(lambda m, dg: m.vLabelCount("C") <= 4)
# Hide the reactions with the large molceules as well:
def dRefEval(dRef):
	der = dRef.derivation
	if any(m.vLabelCount("C") > 4 for m in der.left): return False
	if any(m.vLabelCount("C") > 4 for m in der.right): return False
	return True
p.pushEdgeVisible(dRefEval)
# Add the number of Cs to the molecule labels:
p.pushVertexLabel(lambda m, dg: "\\#C=" + str(m.vLabelCount("C")))
# Highlight the molecules with 4 Cs:
p.pushVertexColour(lambda m, dg: "blue" if m.vLabelCount("C") == 4 else "")
# Print the network with the customised printer.
dg.print(p)
\end{lstlisting}

\end{document}